\documentclass[reprint,amsmath,amssymb,aps]{revtex4-1}

\usepackage{graphicx}
\usepackage{dcolumn}
\usepackage{bm}
\usepackage[utf8]{inputenc}
\usepackage{hyperref}
\usepackage{float}
\usepackage[abs]{overpic}
\usepackage{gensymb}
\usepackage{upgreek}
\usepackage{color}

\makeatletter
\renewcommand\@makecaption[2]{%
	\par
	\vskip\abovecaptionskip
	\begingroup
	\small\rmfamily
	\begingroup
	\samepage
	\flushing
	\let\footnote\@footnotemark@gobble
	\@make@capt@title{#1}{#2}\par
	\endgroup
	\endgroup
	\vskip\belowcaptionskip
}

\begin{document}
	
\preprint{APS/123-QED}
	
	\title{Systematic Raman study of optical phonons in $R$Ba$_2$Cu$_3$O$_{6+\delta}$ ($R$ = Y, Dy, Gd, Sm, Nd): Antiferromagnetic coupling strength versus lattice parameters}
	
\author{Silvia Müllner$^{1}$}
\author{Wayne Crump$^{2}$}
\author{Dirk Wulferding$^{1}$}
\author{\mbox{Benjamin P. P. Mallett$^{3, 4}$}}
\author{Peter Lemmens$^{1}$}
\author{Amit Keren$^{5}$}
\author{Jeffery L. Tallon$^{2, 4}$}

\affiliation{
	\mbox{$^{1}$Institute for Condensed Matter Physics, Technical University of Braunschweig, D-38106 Braunschweig, Germany}
	\mbox{$^{2}$Robinson Research Institute, Victoria University of Wellington, P.O. Box 33436, Lower Hutt 5046, New Zealand}
	\mbox{$^{3}$Physics and Chemical Sciences, Dodd Walls Centre for Photonic and Quantum Technologies, The Photon Factory,} \mbox{University of Auckland, 38 Princes St, Auckland, New Zealand}
	\mbox{$^{4}$MacDiarmid Institute for Advanced Materials and Nanotechnology, School of Chemical and Physical Sciences,} \mbox{Victoria University of
	Wellington, PO Box 600, Wellington, New Zealand}
	\mbox{$^{5}$Department of Physics, Technion–Israel Institute of Technology, Haifa 32000, Israel}}
\date{\today}

\begin{abstract}	
	We present a systematic study of the interplay between lattice parameters and the energy of the optical phonons as well as the antiferromagnetic coupling strength, $J$, in the high-$T_{\text{c}}$ superconducting cuprate  $R$Ba$_2$Cu$_3$O$_{6+\delta}$ (\mbox{$R$-123}, $R=$ Y, Dy, Gd, Sm, Nd) with hole doping $p$ ($0.00<p\lesssim0.04$). The energy of the $B_{1g}$ mode at \mbox{$\upnu_{B1g}\approx335$~cm$^{-1}$} has been found to relate systematically to the inverse of the lattice parameter $a$. Our results confirm the temperature dependent phonon splitting for \mbox{Nd-123} at low doping, which has been reported for optimally doped \mbox{Nd-123}. Surprisingly, $J$ is independent of $a$ for the first four $R$ families, and a general consistency between $T_{c}^{\text{max}}$ and $J$, as suggested in a previous investigation, could not be confirmed. 
\end{abstract}	
	\maketitle
		
	\section{Introduction}
	$R$Ba$_2$Cu$_3$O$_{6+\delta}$ \mbox{($R$-123)} hosts a variety of electronic and magnetic properties. The mechanisms leading to the complexity of its phase diagram, e.g., antiferromagnetic (AFM), pseudogap, metallic, superconducting states are still unresolved \cite{quantummatter,SC}. Its physical properties depend on the temperature, and doping, which is controlled by the oxygen content $\delta$ \cite{charge, Jorgensen}. At optimal hole \mbox{doping} \mbox{($p\approx0.16$)}, it reaches a superconducting transition temperature of \mbox{$T_{c}^{\text{max}}\approx100$~K}  \cite{TcmaxY,TcmaxNd}. At low doping \mbox{($p\lesssim0.05$)} it is an AFM Mott insulator \cite{Mott,mottinsulator} with an AFM coupling strength on the order of \mbox{$J\approx100$~meV} \cite{J,ev}. The Heisenberg model describes the long-range AFM ordering via nearest-neighbor interaction of the spin carriers but does not explain the doping dependence. The generally accepted Hubbard model deals with the doped case although it fails to explain the material dependent systematics, for example, the properties of the refractive sum \cite{shannon}. The Heisenberg and Hubbard models in their simplest form are related at strong coupling via $J\propto t^2/U$, where $t$ is the hopping parameter and $U$ the on-site interaction.
	
	Our study focuses on the electronic and magnetic properties of \mbox{$R$-123} by analyzing the phonon modes and two-magnon scattering from Raman scattering experiments. The required energy for a spin-flip process depends on $J$ and relates to the Raman shift of the two-magnon peak, $\upnu_{\text{2m}}$ \cite{Parkinson,conversiontoJ, Chubukov,Weidinger}. Since $U$ is not expected to vary between materials, measurements of $J$ actually reflect on $t$, a parameter that is relevant at all doping.
	
	The lattice parameters are controlled by the radius of the rare-earth ($R$) ion, $r_R$. Our \mbox{$R$-123} samples ($R=$ Y, Dy, Gd, Sm, Nd) order antiferromagnetically. The magnetic moment resides on the Cu ion, which lie within the crystallographic $ab$ plane. These samples have a hole-doping content of $p~(0.00 < p \lesssim 0.04)$ (see Table~\ref{tab:S}) in which the lattice parameters $a$ and $b$ are equal \cite{spacegroup,space}. Reference values of $r_R$ and the lattice parameter $a$ and $c$ were obtained from a neutron diffraction study by Guillaume \textit{et al.} \cite{Guillaume} and are found in Table \ref{tab:lattice}. 
	\section{Experimental}
	Long-range magnetic order decreases with increasing temperature; hence the two-magnon data was collected at \mbox{$T=15$~K}. A helium flow cryostat (KONTI-cryostat-Mikro) cooled the samples in a vacuum environment. The micro-Raman setup (Jobin-Yvon, LabRAM HR) operates with a liquid nitrogen cooled CCD camera (Horiba, Spectrum One) and a Nd:YAG solid-state laser which emits an excitation wavelength of $\lambda_{\text{exc}}=532.1$~nm. The confocal setup with 50x magnification produced a laserspot on the sample with a diameter of $d_\text{spot}\approx10~\upmu$m and a laser power of $P_\text{laser}\approx150~\upmu$W. We use the Porto notation to describe the light polarization and orientation of the sample. The low temperature measurements were taken in \mbox{$-z(x^\prime y^\prime)z$} backscattering geometry because two-magnon scattering is strongest in \mbox{$B_{1g}$ symmetry} \cite{Chelwani}.
	
    Each sample is an agglomerate of single crystals of sizes up to $l_\text{crystal}\approx100~\upmu$m. The doping of the crystals has been adjusted by annealing in argon, typically at $T \approx650~^\circ$C, followed by quenching into liquid nitrogen. In each case, the doping state was determined using measurements of thermoelectric power \cite{Obertelli} (i.e., Seebeck coefficient) at room temperature, \mbox{$S_{290} = 372$~exp($-32.4p$) \cite{Sconvertedtop}}, and confirmed to be close to zero hole doping. Since the two-magnon Raman shift can be analyzed in terms of the Heisenberg model only in the undoped case, the lightly doped samples allow us to either verify that our results are doping independent or extrapolate them to zero doping if needed.
\begin{table}[H]
	\caption{\label{tab:S}Our investigated \mbox{$R$-123} samples.}		
	\begin{ruledtabular}
		\begin{tabular}{lc}
			$R$-123 &  $p$ (holes per Cu atom) 			\\
			\hline
			Y-123  & 0.027, 0.017, 0.000				\\
			Dy-123 & 0.026, 0.023, 0.020, 0.001, 0.000	\\
			Gd-123 & 0.024, 0.017, 0.003, 0.001, 0.000	\\ 
			Sm-123 & 0.024, 0.020, 0.006, 0.000			\\
			Nd-123 & 0.036, 0.025, 0.000					\\ 
		\end{tabular}
	\end{ruledtabular}  
\end{table}		
\section{Results and discussion}
\begin{figure}
	\includegraphics[width=87mm]{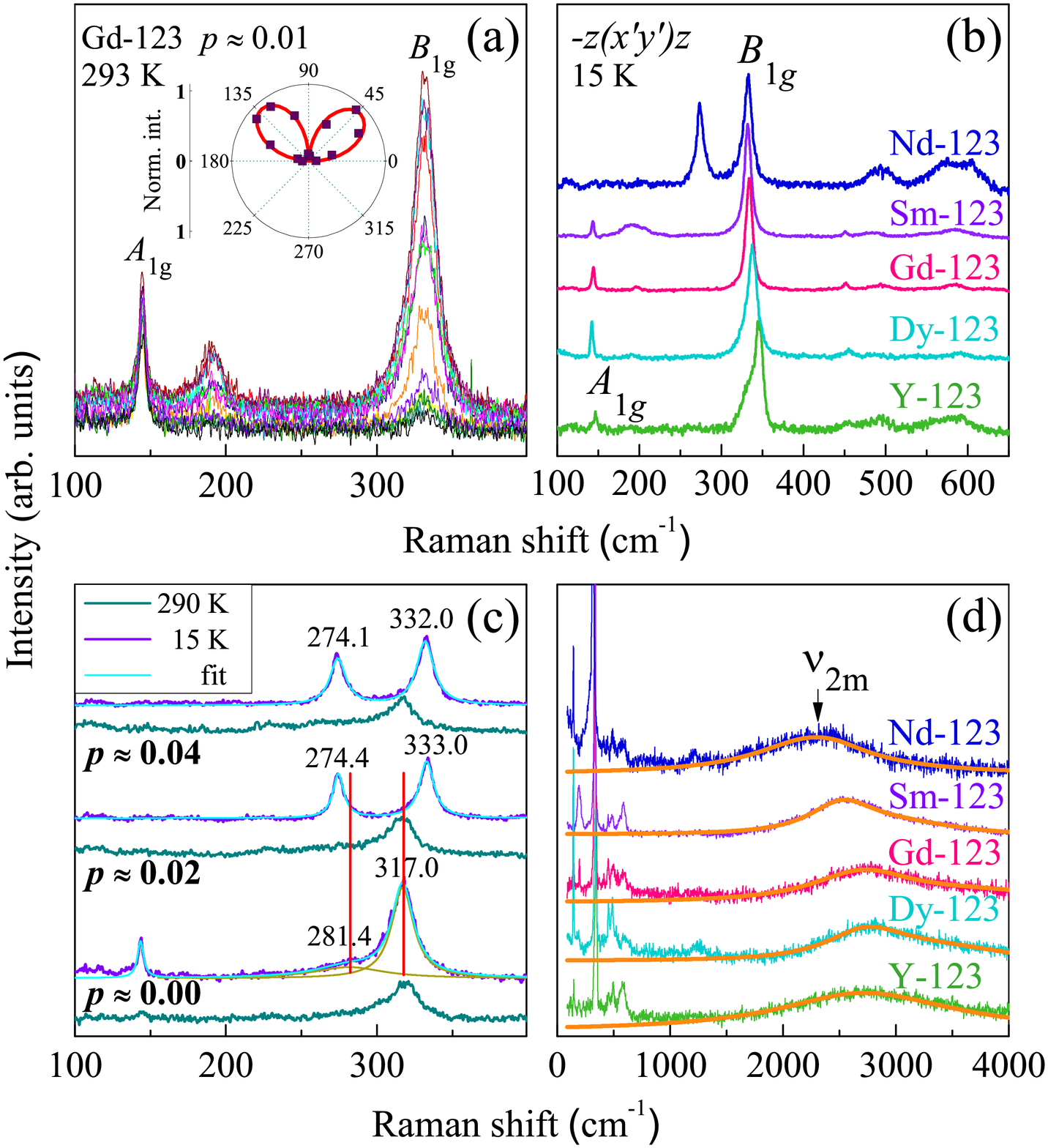}
	\caption{\label{fig:Ramanspectra}(a) Raman spectra of \mbox{Gd-123} in different light polarizations. The integrated intensity of the \mbox{$B_{1g}$ mode} is fitted with a sine function (inset). (b) Phonon spectra of different \mbox{$R$-123}. (c) \mbox{Nd-123} at different $p$ and $T$. (d) Two-magnon excitation at $\upnu_{\text{2m}}$ from different \mbox{$R$-123}, fitted with a Lorentzian and an additional Gaussian function.}
\end{figure}

	The orientation of the crystal has been determined by polarization dependent measurements at room temperature, in which each sample was rotated in steps of approximately $15^\circ$ and measured sequentially; see \mbox{Fig.~\ref{fig:Ramanspectra}(a)}. The highest scattering intensity of the mode at $\upnu_{B1g}\approx335$~cm$^{-1}$ determines the \mbox{$B_{1g}$ geometry}, whilst the low intensity of this mode at $0^{\circ}$ (and symmetry related angles) indicates good crystallinity within the measurement area. Figure~\ref{fig:Ramanspectra}(b) shows the phonon spectra of different \mbox{$R$-123}. Each hosts a mode at \mbox{$\upnu_{B1g}\approx335$~cm$^{-1}$} involving antiphase vibrations of the oxygen ions in the CuO$_2$ layers. A broad mode with weak intensity emerges at \mbox{$\upnu\approx198$~cm$^{-1}$} in \mbox{Sm-123} and also in \mbox{Gd-123}, \mbox{Dy-123}, and \mbox{Y-123}. This mode has been observed in previous studies on AFM \mbox{Y-123} \cite{broadmode}.  Our polarization dependent measurement, see Fig.~\ref{fig:Ramanspectra}(a), shows that this mode has \mbox{$B_{1g}$ character}.	
	
	Some additional modes appear around \mbox{$\upnu\approx450$ cm$^{-1}$}, $500$ cm$^{-1}$, and \mbox{$600$ cm$^{-1}$} and are associated with, respectively, in-phase O vibration on the CuO$ _{2} $ sites, O vibration on the apical site, and O vibration on the \mbox{Cu-O} chain sites \cite{modes,Limonov}. 	
    The \mbox{$A_{1g}$ mode} at $\upnu_{A1g}\approx 145$~cm$^{-1}$ involves vibrations of the in-plane Cu ions \cite{Inder}. In \mbox{Nd-123}, this mode is only observed at $p\approx0.00$ [see \mbox{Fig.~\ref{fig:Ramanspectra}(c)}]. The absence of the \mbox{$A_{1g}$ mode} with even slightly higher $p$ correlates with the phonon splitting of the  \mbox{$B_{1g}$ mode} in \mbox{Nd-123}. This phonon splitting has been observed in optimally doped \mbox{Nd-123} \cite{ionsize,crystal-field,crystal-field2}. It is ascribed to the coupling and mixing of the $B_{1g}$ mode to a crystal-field excitation that involves a Nd$^{3+}$ 4$f$ electron. Our data confirms this splitting between \mbox{$(0.02 \lesssim p \lesssim 0.04)$} in which the modes have a Raman shift of \mbox{$\upnu_{B1g}\approx332$~cm$^{-1}$} and \mbox{$\upnu_{\text{ph}}\approx274$~cm$^{-1}$}. No such splitting is observed at $p\approx0.00$ and all temperatures. With increasing $p$, the structure of \mbox{Nd-123} might become more susceptible to changes and phonon-crystal-field splitting. An anomalous shift of the phonon crystal-field excitation at room temperature is observed throughout the investigated doping region, i.e., the phonon has a Raman shift of $\upnu_{B1g}\approx317$~cm$^{-1}$ instead of $\upnu_{B1g}\approx332$~cm$^{-1}$ as illustrated in Fig.~\ref{fig:Ramanspectra}(c). 
	
	Figure~\ref{fig:Ramanspectra}(d) depicts the two-magnon spectra with the peak position at $\upnu_{\text{2m}}$. For spin $S=1/2$ systems, as present in the examined samples, $J$ relates to  $\upnu_{\text{2m}}$ with $J=\upnu_{2m}/3.22$ \cite{Parkinson,conversiontoJ}, $J=\upnu_{2m}/2.8$ \cite{Chubukov}, or \mbox{$J=\upnu_{2m}/2.84$} \cite{Weidinger} depending on the theoretical approach; we chose the first one to be consistent with Ref. \cite{dielectricRaman}. The peaks are predominantly fitted with a Lorentzian function and an additional Gaussian function towards the higher Raman shift region of $\upnu_{\text{2m}}$. We ascribe this to contributions from magnons and electrons, respectively \cite{correlated}. 
	
\begin{figure} [H]
	\includegraphics[width=87mm]{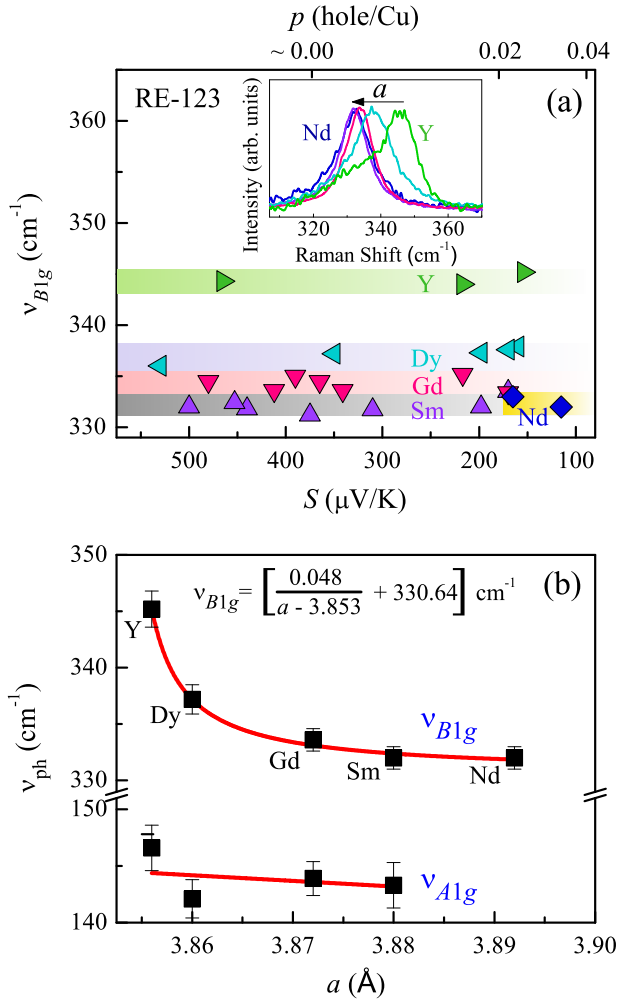}
	\caption{\label{fig:A1gB1g}(Inset) \mbox{$B_{1g}$ mode} of different \mbox{$R$-123.} (a) Raman shift of the \mbox{$B_{1g}$ mode} plotted versus the doping $p$, which is associated with $S$. (b) Average Raman shift of the $A_{1g}$ and \mbox{$B_{1g}$ modes} for the different \mbox{$R$-123} compared with the lattice parameter $a$. The \mbox{$B_{1g}$ mode} is fitted with the indicated function and the \mbox{$A_{1g}$ mode} has a linear fit.}
\end{figure}

	Figure~\ref{fig:A1gB1g}(a) shows the Raman shift of the \mbox{$B_{1g}$ mode}, $\upnu_{B1g}$. The data points are assigned to a certain symbol and color, corresponding to the $R$ ion. The Raman shift of the \mbox{$B_{1g}$ mode} remains unaffected within our investigated doping region, illustrated with the shaded areas. The inset of Fig.~\ref{fig:A1gB1g}(a) depicts  spectra of the \mbox{$B_{1g}$ mode} from \mbox{Nd-123,} \mbox{Sm-123,} \mbox{Gd-123,} \mbox{Dy-123,} and \mbox{Y-123} (left to right). The Raman shift of the two modes, $\upnu_{B1g}$ and $\upnu_{A1g}$, are listed in Table~\ref{tab:lattice} and are plotted versus the lattice parameter $a$ in Fig.~\ref{fig:A1gB1g}(b). The Raman shift of the \mbox{$B_{1g}$ mode} monotonically decreases with increasing lattice parameter $a$ in agreement with previous studies \cite{FriedlPRL90}. We find that it relates to $a$ according to
	\begin{equation*}
		\upnu_{B1g}=\bigg[\frac{0.048}{a-3.853}+330.64\bigg]~\text{cm$^{-1}$.} 
	\end{equation*}
	The \mbox{$A_{1g}$ mode} has a similar trend with a larger margin of error due to the weak scattering intensity of this mode. 
	
	The analysis seen in Fig.~\ref{fig:A1gB1g}(b) demonstrates that the Raman shift of the \mbox{$B_{1g}$ mode} undergoes a greater change with $a$ than the \mbox{$A_{1g}$ mode}. A possible explanation for this might be the proximity of the modes to the $R$ ion, because the radius of the $R$ ion is varied to tune the lattice parameters. The \mbox{$B_{1g}$ mode} is about half the distance to the $R$ ion than the \mbox{$A_{1g}$ mode}. Also noteworthy is the lower limit of the \mbox{$B_{\text{1g}}$ mode} at \mbox{$\upnu_{\text{B1g,\text{~min}}}\approx331$~cm$^{-1}$} towards larger $a$. 	
\begin{table*}
	\caption{\label{tab:lattice}Literature values of $r_R$, $a$, $b$, $c$, \mbox{$T_{c}^{\text{max}}$} and buckling angle on CuO$_2$ plane of \mbox{$R$-123} for $p<0.05$ \cite{Guillaume,TcmaxY,TcmaxNd,dielectricRaman}.
	Our data is indicated with an asterisk. Our $\upnu_{\text{2m}}$ and some reference values of $\upnu_{\text{2m}}$ \cite{dielectricRaman,Pr} are converted to $J$ \cite{Parkinson}.}  		
	\begin{ruledtabular}
		\begin{tabular}{lcccccccccl}
			$R$-123 & $r_R$ & $a$, $b$ & c & $T_{c}^{\text{max}}$ & buckling angle  & $\upnu_{\textit{A}{1g}}^*$ & $\upnu_{\textit{B}{1g}}^*$ & $\upnu_{\text{2m}}^*$ & reference  $\upnu_{\text{2m}}$ & ~$J$ \\
			& (\AA)    & (\AA) & (\AA) &(K)  & a/b-axis ($^{\circ}$) & (cm$^{-1}$) & (cm$^{-1}$) & (cm$^{-1}$)  & (cm$^{-1}$) & (meV)\\
			\hline
			Y-123  & 1.019 & 3.856 & 11.793 & 93.5 & 6.2  & 146.6 & 345.2 & 2634 & 2615 & ~101.5$^*$\\
			Dy-123 & 1.027 & 3.860 & 11.796 & 92.2 & 6.04 & 142.1 & 337.2 & 2749 & 2678 & ~105.8$^*$\\
			Gd-123 & 1.053 & 3.872 & 11.807 & 93.6 & 6.06 & 143.9 & 333.6 & 2735 & 2620  & ~105.3$^*$\\
			Eu-123 & 1.066 & 3.879 & 11.811 & 94.0 & 5.43 &  n.a  & n.a   & n.a. & 2610 & ~100.5 \\
			Sm-123 & 1.079 & 3.880 & 11.815 & 94.7 & 5.63 & 143.3 & 332.0 & 2574 & 2605 & ~~99.1$^*$  \\
			Nd-123 & 1.109 & 3.893 & 11.830 & 96.1 & 6.90 & n.a.  & 332.1 & 2294 & 2525 &  ~~88.3$^*$   \\
			Pr-123 & 1.126 & 3.900 & 11.832 &  n.a & 6.06 &  n.a  & n.a.  & n.a. & 2190 & ~~84.3 \\
		\end{tabular}
	\end{ruledtabular}
\end{table*} 	
	
	\begin{figure}
		\includegraphics[width=87mm]{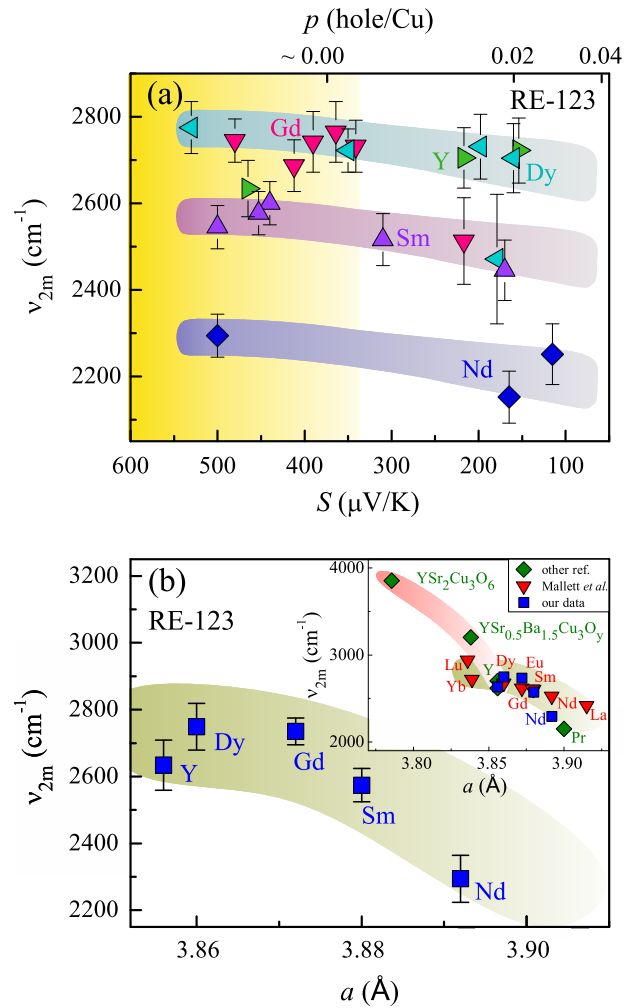}
		\caption{\label{fig:J-vs-latticeparameters}(a) Two-magnon peak, $\upnu_{2m}$, versus doping, $p$, and thermoelectric power, $S$. The yellow region indicates \mbox{$p\approx0.00$} and the shaded areas illustrate the trend towards lower $\upnu_{2m}$ with increasing $p$, in e.g., \mbox{Dy-123}, \mbox{Sm-123}, and \mbox{Nd-123}. (b) Average values of $\upnu_{\text{2m}}$ of each \mbox{$R$-123} taken from the yellow shaded area in (a) and plotted versus $a$. (Inset) Our data (blue squares) and other data (red triangles and green diamonds) to cover the widest possible data range \cite{YSr2,dielectricRaman}.}
	\end{figure}
	As seen in Fig.~\ref{fig:Ramanspectra}(d), the Raman shift of $\upnu_{\text{2m}}$ is only marginally affected within the investigated doping range. Generally, the two-magnon peak hardens with decreasing doping and is consistent with previous studies on other \mbox{high-$T_{\text{c}}$} SC cuprates \cite{Dirk,Sugai,Brenig}). This trend is illustrated with the shaded areas in Fig.~\ref{fig:J-vs-latticeparameters}(a) for \mbox{Dy-123}, Sm-213 and \mbox{Nd-123}. The yellow shaded area indicates the region for $p\approx0.00$ ($S~\gtrsim~340~\upmu$V/K). From this region, the average value of $\upnu_{\text{2m}}$ for each \mbox{$R$-123} is plotted versus $a$ in Fig.~\ref{fig:J-vs-latticeparameters}(b) and is listed in Table~\ref{tab:lattice} as well. The olive green shaded area shows the systematic relation between $\upnu_{2m}$ and $a$. The surprising result is that from \mbox{Y-123} to \mbox{Sm-123} there is no change in $\upnu_{\text{2m}}$ within experimental errors and only for \mbox{Nd-123} a downward trend is observed. This stands in contrast to the systematic behavior of the \mbox{$B_{1g}$ mode}. $J$ is independent of $a$ for four out of five samples we reexamined.
	
	Reference values of $\upnu_{\text{2m}}$ for YSr$_2$Cu$_3$O$_6$ \cite{YSr2}, YSr$_{0.5}$Ba$_{1.5}$Cu$_3$O$_6$ \cite{ionsize,dielectricRaman}, and other lanthanide (Ln) \mbox{Ln-123} are considered in the inset of Fig.~\ref{fig:J-vs-latticeparameters}(b) shown with green diamonds and red triangles \cite{dielectricRaman} in order to cover the widest possible range for $J$ versus $a$. These compounds have the same structure as \mbox{$R$-123.} However, YSr$_2$Cu$_3$O$_6$ is synthesized under high oxygen pressure \cite{YSr2,YSr2-2} and, along with YSr$_{0.5}$Ba$_{1.5}$Cu$_3$O$_6$, is more compressed because Sr is substantially smaller than Ba while the compounds are otherwise isostructural with \mbox{Y-123}.
	Nevertheless, the different out-of-plane structure  possibly renders these last two compounds in a class on their own. The green and pink shaded areas in the inset of Fig.~\ref{fig:J-vs-latticeparameters}(b) emphasize these two classes yet together they show a systematic decrease of $\upnu_{\text{2m}}$ with increasing $a$ for \mbox{$R$-123}, but perhaps with different rates. 
	As \mbox{Y-123} is not part of the lanthanide series plus the fact that its orbital structure differs slightly from the \mbox{Ln-123,} it is not clear to which class it belongs (green or pink). Hence it is represented in both areas. The uniqueness of Y might be the reason for the different $a$ dependence of $\upnu_{\text{2m}}$ in the pink area compared with the remaining \mbox{Ln-123}. The apical oxygen bond length could be a crucial factor that influences $J$ \cite{apical} and might account for the low value of $J$ in the \mbox{Y-123} compared with the other Y based compounds. Beyond that we need to recognize that any small variations may be attributable to compositional or doping variations, particularly noting that Raman and thermopower effectively probe different depths in samples where we have attempted, with difficulty, to remove the last dopant oxygen.
	
\begin{figure}[H]
	\includegraphics[width=87mm]{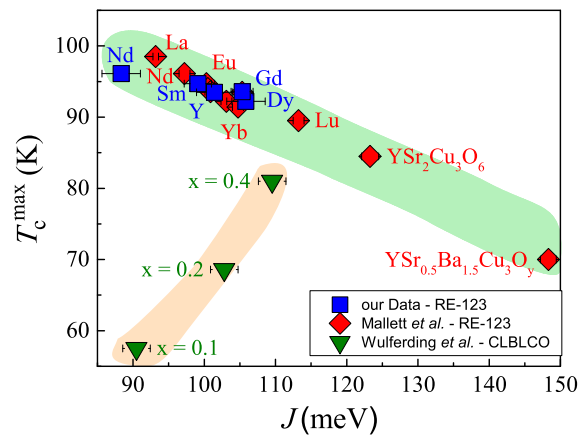}
	\caption{\label{fig:TcJ}Maximum transition temperature $T_{c}^{\text{max}}$ as a function of the AFM coupling strength, $J$ \cite{Parkinson, Chubukov,Weidinger}, of \mbox{$R$-123} in the green shaded area. The red diamonds and blue squares are data from Mallett \textit{et al.} \cite{dielectricRaman} and us, respectively. The YSr$_2$Cu$_3$O$_6$ and YSr$_{0.5}$Ba$_{1.5}$Cu$_3O$$_6$ were synthesized under high pressure \cite{YSr2,YSr2-2}. The green triangle shows data of an isostructural compound to $R$,  (Ca$_x$La$_{1-x}$)(Ba$_{1.75-x}$La$_{0.25+x}$)Cu$_3$O$_{y}$ (CLBLCO) from Wulferding \textit{et al.} \cite{Dirk}.}
\end{figure}

    Studies on whether $J$ relates to $T_{c}^{\text{max}}$ have led to conflicting findings. \mbox{Wulferding \textit{et al.}} observed, using Raman, a linear increase of \textit{T}$_{c}^{\text{max}}$ with increasing $J$ in (Ca$_x$La$_{1-x}$)(Ba$_{1.75-x}$La$_{0.25+x}$)Cu$_3$O$_{y}$ (CLBLCO), an isostructural compound to \mbox{$R$-123,} and proposed a correlation between these parameters \cite{Dirk}. Their measurements are in agreement with $J$ determination by muon spin rotation \cite{OferPRB06}, resonance inelastic x-ray scattering \cite{EllisPRB15}, and with $t$ measurement by angle resolved photoemission spectroscopy \cite{DrachuckPRB14}.  On the other hand, \mbox{Mallett} \textit{et al.} observed a decreasing $T_{c}^{\text{max}}$ with $J$ but also found the opposite behavior when applying external pressure on \mbox{$R$-123} \cite{dielectricRaman,Tallon}. Both, the data of Wulferding \textit{et al.} and Mallett \textit{et al.} regarding  $T_{c}^{\text{max}}$ versus $J$ are compared to our data in Fig.~\ref{fig:TcJ}. The data of Mallett \textit{et al.} is in good agreement with our data (apart from the \mbox{Nd-123} sample). The contradicting result of Wulferding \textit{et al.} to the other two results suggests that no general conclusion between $T_{c}^{\text{max}}$ and $J$ can be drawn based on Fig.~\ref{fig:TcJ}. Reasons for this contradiction have been proposed \cite{Tallon}. 
   	\section{Summary}
     A systematic relation between the energy of the optical phonons and the lattice parameter $a$ could be determined. In previous studies a \mbox{$T$-dependent} phonon splitting of the \mbox{$B_{1g}$ mode} was observed in optimally-doped \mbox{Nd-123} \cite{crystal-field,crystal-field2}. Our results confirm this splitting for low hole doping, but not for undoped \mbox{Nd-123}.
     For the samples we reexamined, the AFM coupling strength, $J$, did not decrease monotonically with increasing lattice parameter $a$. When considering other studies, including samples prepared under high pressure, a general trend of decreasing $J$ with increasing $a$ becomes apparent, although with a plateau around $a=3.85$ \AA. A universal relation between $T_{c}^{\text{max}}$ and $J$ as suggested previously \cite{Dirk} could not be established.
\section{Acknowledgements}
   This work has been made possible by the support of the Deutsche Forschungsgemeinschaft DFG Grant No. LE967/16-1 and QUANOMET initiative within Project \mbox{LENA-NL4-1}, and by the German-Israeli Foundation for Scientic Research and Development (GIF). BPPM acknowledges support from the Rutherford Foundation of New Zealand. We thank Kim Paul Schmidt, Dr. Roser Valentí and Bo Liu for constructive discussions.	
	\bibliographystyle{ieeetr}
	\bibliography{biblio}	
	\appendix	
\end{document}